\pdfoutput=1
\documentclass[iop,apjl]{emulateapj}

\usepackage{url}
\usepackage{natbib}
\bibpunct{(}{)}{,}{a}{}{;}
\usepackage{amssymb}
\usepackage{amsmath}
\usepackage{url}
\usepackage{graphicx}

\newcommand{\myemail}{lkristensen@cfa.harvard.edu}

\shorttitle{Velocity-resolved hot water emission detected toward HL Tau with the Submillimeter Array}
\shortauthors{Kristensen et al.}

\begin{document}

\title{Velocity-resolved hot water emission detected toward HL Tau with the Submillimeter Array}

\author{Lars E. Kristensen, Joanna M. Brown, David Wilner}
\affil{Harvard-Smithsonian Center for Astrophysics, 60 Garden Street, Cambridge, MA 02138, USA}
\email{email: \myemail}
\author{Colette Salyk}
\affil{Vassar College, Department of Physics and Astronomy, 124 Raymond Avenue, Poughkeepsie, NY 12604, USA}

\begin{abstract}
Using the Submillimeter Array (SMA) on Mauna Kea, the H$_2^{16}$O 10$_{2,9}$--9$_{3,6}$ transition ($E_{\rm up}$=1863~K) at 321.2~GHz has been detected toward the embedded low-mass protostar HL Tau. The line centroid is blue-shifted by 15~km\,s$^{-1}$ with respect to the source velocity, and it has a $FWHM$ of 20~km\,s$^{-1}$. The emission is tentatively resolved and extends $\sim$3--4$''$ over the sky ($\sim$2~beams), or $\sim$500~AU at the distance of Taurus. The velocity offset, and to a lesser degree the spatial extent of the emission, shows that the line originates in the protostellar jet or wind. This result suggests that at least some water emission observed toward embedded sources, and perhaps also disk sources, with \textit{Herschel} and \textit{Spitzer} contains a wind or jet component, which is crucial for interpreting these data. These pathfinder observations done with the SMA opens a new window to studying the origin of water emission with e.g. ALMA, thus providing new insights into where water is in protostellar systems. 
\end{abstract}


\section{Introduction}

Water plays a key role in the formation of stars and planets. It is one of the dominant coolants of warm and hot molecular gas \citep[300$\lesssim$$T$$\lesssim$1500~K, e.g.,][]{karska13}, for example in star-forming regions. Furthermore, water is an excellent probe of dense shocked gas in embedded protostellar systems \citep[e.g.][]{kristensen13}. During later evolutionary stages, water may play a key role in planet formation \citep{stevenson88} and it may trace the radiation environment and gas density in disks \citep{bethell09}. See \citet{vandishoeck14} for a full review. The main isotopolog of water, H$_2^{16}$O, is difficult to observe from the ground because of the blocking atmosphere, and primarily maser transitions and highly excited rovibrational transitions at infrared wavelengths have been observed so far \citep[e.g.][]{menten91, salyk08, pontoppidan10a}. Space telescopes, such as the \textit{Spitzer} Space Telescope \citep{gallagher03} and \textit{Herschel} Space Observatory \citep{pilbratt10}, made observations of rotationally excited water technically feasible, although these missions all lack the spatial and spectral resolution required to precisely pinpoint the origin of water emission. 

Water emission may serve as an evolutionary tracer of young stellar objects (YSOs). During the embedded stages, water is uniquely associated with shocks and outflow activity and no detected emission is coming from the disk or envelope \citep[e.g][]{herczeg12, karska14}. At later evolutionary stages, the disk appears to dominate the picture: these sources often do not have any detectable outflow activity associated with them, although the sources are still accreting \citep[e.g.][]{carr08, hogerheijde11, riviere12}. Particularly, for disk sources, the hot water detected by both \textit{Spitzer} and \textit{Herschel} seemingly originates in the inner AU of the disk, where it is hot enough that all water is sublimated from the grains, and where the gas is dense enough that the lines are excited \citep[][]{riviere12, howard13, fedele13}. This scheme raises a number of questions, for example, when does water switch from being a shock tracer to a disk tracer? And if the more evolved sources are still accreting (i.e. also ejecting) material, how certain is the interpretation that the hot water is coming from the inner disk and not a shock? 

HL Tau shows characteristics of both deeply embedded YSOs (envelope, molecular outflow) and disk sources (spatially resolved, Keplerian disk) and is clearly intermediate in evolutionary stage between the so-called Class I and II stages \citep[][and references therein]{alma15}. \textit{Herschel}-PACS detected hot H$_2^{16}$O emission ($J_{KaKc}$=8$_{18}$--7$_{07}$, 63.32~$\mu$m, $E_{\rm up}$=1071~K) toward this source \citep{riviere12}. Because of the low spectral and spatial resolution ($\sim$100~km\,s$^{-1}$ and 9\farcs4, respectively), interpreting the origin of emission is not straightforward and requires a model. \citet{riviere12} interpreted the emission as coming from the inner disk ($R$$\sim$1--3~AU), but a wind origin cannot be excluded\footnote{``Wind'' here refers to anything that is launched from or near the protostar, whereas ``outflow'' is reserved for cold ($T\lesssim100$~K) entrained material.}. New observations  of water emission from the system are required to break this degeneracy. 

The H$_2^{16}$O 10$_{2,9}$--9$_{3,6}$ transition at 321.226~GHz ($E_{\rm up}$/$k_{\rm B}$=1863~K) is located on the shoulder of the deep atmospheric H$_2$O absorption feature caused by the 4$_{2,2}$--5$_{1,5}$ transition at 325.1~GHz ($E_{\rm low}$/$k_{\rm B}$=454~K). The 10$_{2,9}$--9$_{3,6}$ transition itself is not affected by atmospheric conditions as the lower-level energy (1846 K) is too high to be excited under normal conditions. However, because of its proximity to the 325~GHz absorption feature, excellent atmospheric conditions are required to observe this water line. The line has previously been observed and detected toward high-mass star-forming regions and AGB stars \citep[e.g.][]{menten91, patel07, hirota14} where it is a well-known maser, but it has not been detected toward low-mass YSOs. 

We here report on the detection of H$_2^{16}$O emission toward the embedded disk source HL Tau. Section 2 contains the observational details and results are presented and analyzed in Section 3. The results are discussed and summarized in Section 4.

\section{Observations}

HL Tau (RA=04$^{\rm h}$31$^{\rm m}$38\fs44, dec=+18$^\circ$13$'$57\farcs6) was observed on Dec. 18, 2013 with the SMA in the compact configuration. The receivers were tuned to the H$_2^{16}$O 10$_{2,9}$--9$_{3,6}$ transition at 321.226~GHz, and the transition was placed in chunk 1 of the upper sideband. The amount of precipitable water vapor in the atmosphere was low, PWV$<$1~mm ($\tau_{\rm 225GHz}$$\sim$0.06), throughout the night. Six antennas were available in the array at the time of observation. The absolute flux was calibrated against observations of Ganymede, and observations of 3c454.3 were used for bandpass calibration. The quasars 3c120 and 0510$+$180 were used as gain calibrators. 

Data were reduced and imaged in CASA 4.1 \citep[Common Astronomy Software Applications;][]{mcmullan07}. Data from the upper IF band ($\sim$323--325~GHz) were too close to the 325.1~GHz water absorption feature and were flagged because the $S/N$ was too low on all calibrators. The data reduction followed the standard steps of calibrating the bandpass, followed by calibrating the phase and finally the absolute flux. The continuum was imaged first and HL Tau appears as a point source in the SMA compact configuration, consistent with recent ALMA observations at similar frequencies \citep{alma15}. Because of its high continuum flux density ($\sim$2~Jy\,beam$^{-1}$) the data were self-calibrated and the solution was mapped onto the line channels. After calibration, the continuum was subtracted from the complex visibilities. The line data were imaged with a natural weighting to maximize $S/N$ at the expense of spatial resolution. The resulting beam size is 3\farcs0$\times$2\farcs3 (PA=62$^\circ$). 

\begin{figure}[t!]
\epsscale{1.2}
\plotone{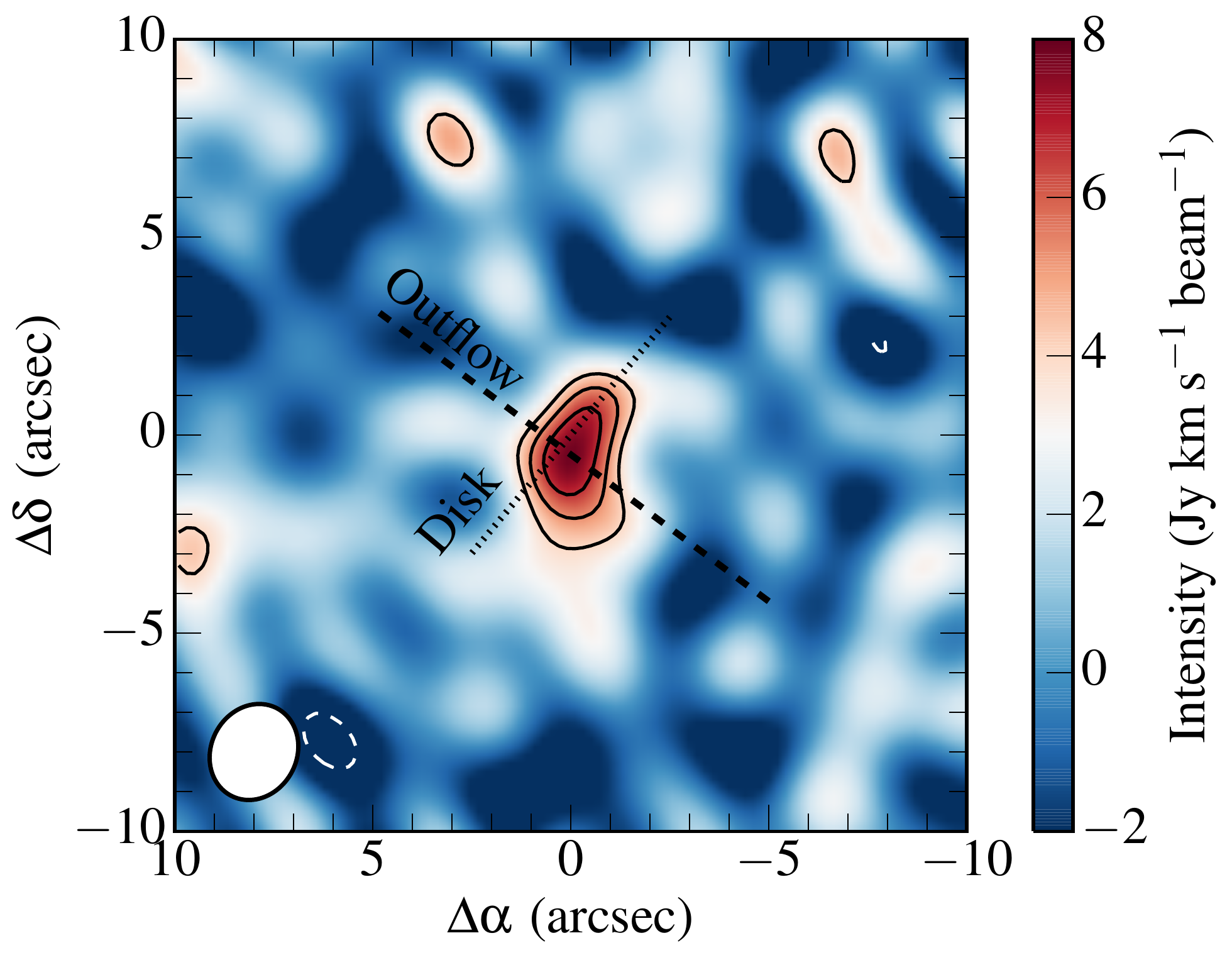}
\caption{H$_2^{16}$O 10$_{2,9}$--9$_{3,6}$ emission toward HL Tau (colors). Black contours show the 3, 4, 5, 6$\sigma$ levels ($\sigma$=1.2~Jy\,km\,s$^{-1}$\,beam$^{-1}$). The beam is marked in white. Emission is integrated over --22.5 to 10~km\,s$^{-1}$  ($v_{\rm source}$=7.0~km\,s$^{-1}$) in the $uv$ plane before imaging; no redshifted emission is detected. The dashed line indicates the direction of the H$_2$ jet and the dotted line shows the direction of the disk. \label{fig:map}}
\end{figure}

\section{Results and analysis}

H$_2^{16}$O 10$_{2,9}$--9$_{3,6}$ is detected towards HL Tau, and a map of integrated emission is shown in Fig. \ref{fig:map}, where emission is integrated from --22.5 to $+$10~km\,s$^{-1}$. The noise level in the map is 1.2~Jy\,km\,s$^{-1}$\,beam$^{-1}$ when measured in emission-free regions\footnote{All quoted uncertainties are rms uncertainties and do not include systematic uncertainties, such as the absolute calibration uncertainty which is estimated to be $\sim$20\%.}. Emission is tentatively extended to the south of the source over a region of $\sim$3--4$''$ or $\sim$500~AU at the distance of Taurus \citep[140~pc; e.g.][]{torres09}, where the extended emission is at the 3--4$\sigma$ level. The extended emission falls between the directions of the disk and outflow \citep{carrasco09}; because of the low significance of this southern extension, we exclude it from further analysis. Observations with better sensitivity are needed to confirm the reality of any spatial extent beyond the size of the SMA beam.

\begin{figure}[t!]
\epsscale{1.4}
\plotone{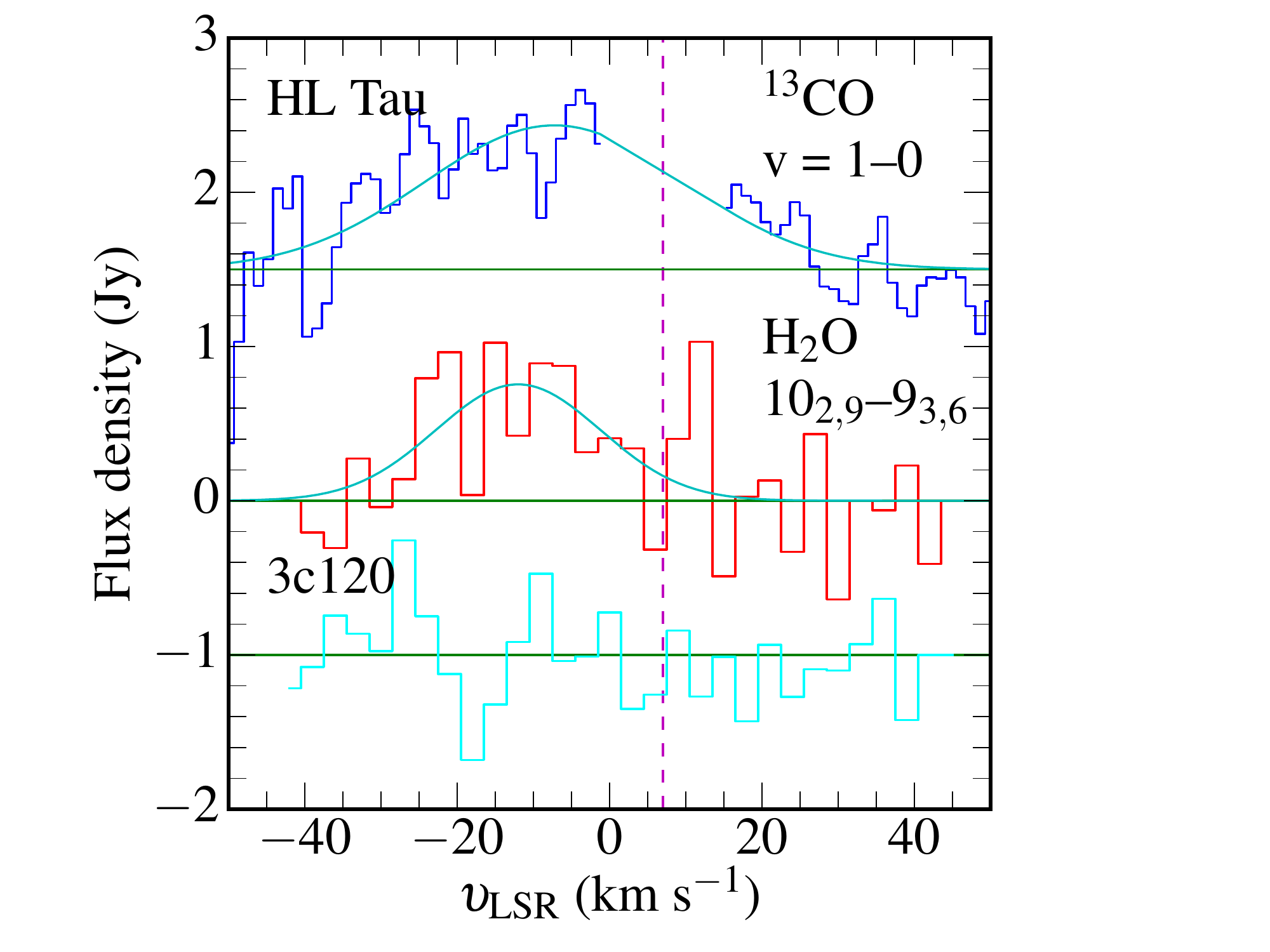}
\caption{H$_2$O 10$_{2,9}$--9$_{3,6}$ spectrum toward HL Tau (red) and co-added $^{13}$CO $v$=1--0 (4.7 $\mu$m) line emission (blue) toward the same source \citep{herczeg11}. The calibrator spectrum of 3c120 is shown for comparison (cyan). Gaussians are fitted to the HL Tau line profiles (cyan), and the baselines are shown (green); the source velocity is marked in magenta. The $^{13}$CO spectrum has been offset for clarity, the units are arbitrary for this spectrum, and the deep envelope self-absorption is masked out. The H$_2$O spectra for both HL Tau and 3c120 are obtained over the same region where emission $>$3$\sigma$ and $r$$<$5$''$ (Fig. \ref{fig:map}). \label{fig:spectra}}
\end{figure}

The spectral profile extracted over the emitting region shows that the line profile is broad, $FWHM$=25$\pm$6~km\,s$^{-1}$, and blue-shifted by 19$\pm$3~km\,s$^{-1}$ from the source velocity \citep[$v_{\rm lsr}$=7.0$\pm$0.2~km\,s$^{-1}$ based on ALMA observations of a number of high-density gas tracers;][]{alma15} (Table \ref{tab:gauss}; Fig. \ref{fig:spectra}). The noise level in the spectrum, when measured over emission-free channels ($>$15~km\,s$^{-1}$) is 0.4~Jy in 3~km\,s$^{-1}$ channels. The integrated intensity over the entire emitting region is 18$\pm$3~Jy\,km\,s$^{-1}$, corresponding to a 6$\sigma$ detection. 

To verify that the line is not an artifact of the bandpass calibration, the complex gain calibrators were imaged in the same way as HL Tau. The emission feature is not present toward either calibrator (see spectrum of the strongest gain calibrator 3c120; Fig. \ref{fig:spectra}) and we conclude the emission is not reduction-related. Furthermore, when plotting just the vector-averaged $uv$ spectrum before cleaning, the line is detected at the same significance as in the cleaned image, and therefore the line is not introduced as an image artifact. 

Because of the apparent offset in velocity, the Splatalogue database\footnote{\url{http://splatalogue.net}} was checked for other potential sources of emission. No other likely candidate is present. The $^{13}$CH$_3$OH 10$_{5,6}$--11$_{4,7}$ transition at 321.203~GHz ($E_{\rm up}$=260~K) falls close to the source velocity, but $^{13}$CH$_3$OH has not been detected toward any source similar to HL Tau, and so we exclude this remote possibility. 

The 321~GHz water transition has previously been detected toward high-mass star-forming regions and AGB stars \citep{menten90, patel07, hirota14}, but never toward a low-mass YSO. The previous detections are, to the best of our knowledge, all masering where the brightness temperature is well in excess of 200~K. The brightness temperature is $\sim$10~K toward HL Tau, but since the emitting region is only very tentatively resolved, the actual brightness temperature may be much higher. However, the large line width ($FWHM$$\sim$25~km\,s$^{-1}$) suggests that emission is thermal rather than masing, as the latter typically show line widths of $<$1--2~km\,s$^{-1}$ \citep[e.g.][]{patel07}. 

With interferometric observations, signals with large spatial extent (specifically, larger than that probed by the shortest baselines) are filtered out. In this case, the emission from the excited H$_2$O line observed with PACS is confined to the central pixel \citep{riviere12}, measuring 9\farcs4$\times$9\farcs4, and so it is unlikely that the even higher-excited water line presented here should be extended beyond the PACS spaxel. In addition, in the SMA compact configuration, spatial scales from $\sim$ 2--20$''$ are recovered and spatial filtering is therefore not an issue. 

A similar line profile is observed in co-added rovibrational $^{13}$CO transitions at 4.7 $\mu$m \citep[$E_{\rm up}$$\gtrsim$3000~K]{herczeg11}, see Table \ref{tab:gauss}; Fig. \ref{fig:spectra}, where the CO emission is compact and spatially unresolved at 0\farcs2 resolution ($\sim$30~AU). The authors attribute the lack of extended emission to the high critical density of the CO transitions, $>$5$\times$10$^{12}$~cm$^{-3}$. For comparison, the critical density of the H$_2$O transition is only 10$^4$~cm$^{-3}$ and so it is likely that the vibrationally excited CO is confined to a comparatively small region near the protostar. 

\begin{deluxetable}{l r r c}
\tabletypesize{\scriptsize}
\tablecaption{Gaussian fits to the H$_2$O and rovibrational CO profiles\tablenotemark{a}. \label{tab:gauss}}
\tablewidth{0pt}
\tablehead{\colhead{} & \colhead{$FWHM$} & \colhead{$v_{\rm LSR}$} & \colhead{$I$} \\
\colhead{} & \colhead{(km s$^{-1}$)} & \colhead{(km s$^{-1}$)} & \colhead{(Jy km s$^{-1}$)}
}
\startdata
H$_2$O 10$_{2,9}$--9$_{3,6}$ & 25(6) & --12(3) & 18(3) \\
$^{13}$CO, $v$ = 1--0\tablenotemark{b} & 40(4) & --7(6) &  
\enddata
\tablenotetext{a}{Values in brackets are 1$\sigma$ uncertainties from the fit.}
\tablenotetext{b}{From \citet{herczeg11}.}
\end{deluxetable}

\subsection{Comparison to radiative transfer models}

A good test to assess the origin of the 321~GHz emission, as well as assessing if the emission arises in the same region as the 63~$\mu$m emission, is to estimate the local excitation conditions of the two lines. In the following, this is done in two ways: first assuming local thermodynamic equilibrium (LTE), second using the non-LTE isothermal code RADEX. 

\citet{carr11} and \cite{salyk11} took a similar approach for analyzing the highly excited water lines observed toward a large sample of classical T Tauri stars with \textit{Spitzer}. Under the assumption of LTE, they found that emission typically originates in warm gas with $T\sim$5--600~K with a high column density (10$^{18}$~cm$^{-2}$) and small emitting radius (1~AU). Modeling of the 63 um water emission line with the disk chemistry code ProDiMo \citep{woitke09}, as well as comparison of line fluxes with Spitzer-based LTE models, suggests this line originates from a slightly larger emitting region with radius 3 AU \citep{riviere12}.

The high flux of the 321~GHz line excludes this line and the 63~$\mu$m emission being in LTE; the value of $N_{\rm up}$/$g_{\rm up}$ is higher for the 321~GHz line than the 63~$\mu$m line, in spite of $E_{\rm up}$ also being higher. The critical density of water lines is typically so high, compared to the ambient density, that the levels are sub-thermally populated. In this case, the critical densities are $\sim$10$^4$~cm$^{-3}$ and 10$^{10}$~cm$^{-3}$ for the 321~GHz and 63~$\mu$m transitions, respectively, at 500~K. Thus, it is possible that the 321~GHz transition is more efficiently excited, in spite of the higher level energy (1800 vs. 1070~K). 

\begin{figure}[t!]
\epsscale{1.2}
\plotone{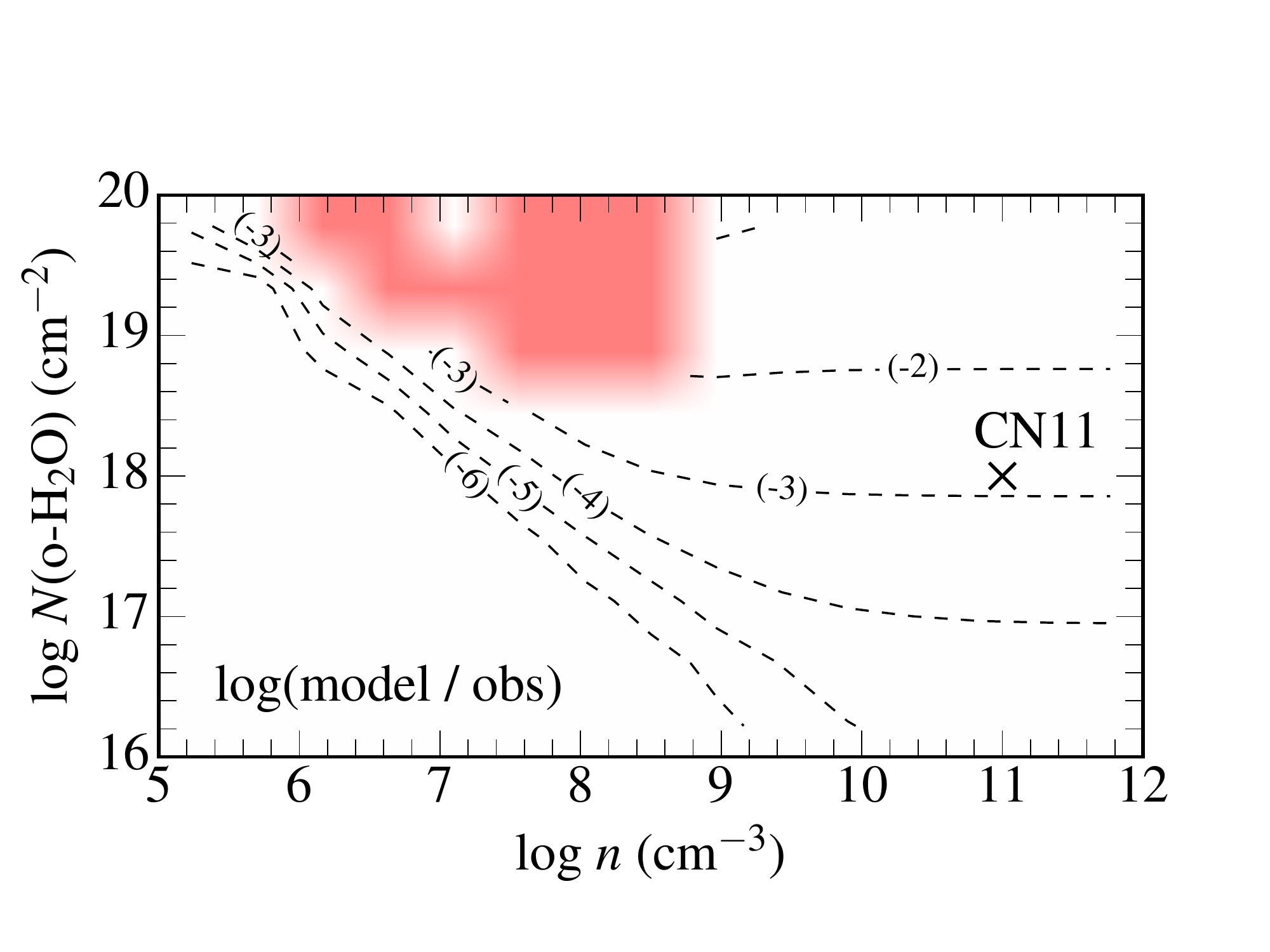}
\caption{Logarithmic model/observed intensity ratios for the H$_2$O 10$_{29}$--9$_{36}$ (321~GHz) / 8$_{18}$--7$_{07}$ (63~$\mu$m) lines as a function of H$_2$ and o-H$_2$O densities. Contours are logarithmically spaced where the zero contour, were it present, would correspond to the observed line ratio. The red region is where the 321~GHz line is masing. The temperature is set to 600~K, similar to that used by \citet{carr11}. The cross marks the \citet{carr11} LTE solution. \label{fig:exc_cond}}
\end{figure}

To test if a non-LTE, but isothermal, solution exists for the observed fluxes, the non-LTE radiative transfer code RADEX is used \citep{vandertak07} to attempt reproducing the observed line ratio of (3.3$\pm$0.6)$\times$10$^{-3}$. The collisional rate coefficients calculated by \citet{faure08} are used in both cases because these include highly excited rotational states as well as rovibrational states; similar results are obtained if the collisional rates from \citet{daniel11} are used. A constant line width of 25~km\,s$^{-1}$ is used in these calculations, and a grid spanning $N$(o-H$_2$O)=10$^{13}$--10$^{20}$~cm$^{-2}$, $n$(H$_2$)=10$^5$--10$^{12}$~cm$^{-3}$, and $T$=200--5000~K was calculated. The observed flux ratio is not found anywhere in this large grid. Figure \ref{fig:exc_cond} shows an example of modeled line ratios vs. the observed one, and the models fall short by orders of magnitude in this grid. There are three likely explanations for this discrepancy: \textit{(i)} the model is inadequate, \textit{(ii)} the 321~GHz line is masering, or \textit{(iii)} the lines do not share a common physical origin. The actual reason may of course also be a combination of these. Each explanation is detailed below. 

The model may be inadequate. The line opacities calculated by the model are very high, often $\gg$100 at high column densities, and the model assumptions no longer apply. Naturally, this mostly applies to the high-column-density regime where a solution, if it exists, is most likely to be found. 

The non-LTE code takes masering into account when calculating the level populations and resulting line intensities, but does not include amplification along the masering path. Models with $N$(o-H$_2$O)$\geq$10$^{19}$~cm$^{-2}$ and $n$(H$_2$)=10$^6$--10$^9$~cm$^{-3}$ showed negative line opacities, $\tau_{\rm 321\,GHz}$$<$--1, and are thus masing. These values are consistent with an analysis based on the masering constraints presented by \citet{neufeld90}. However, the large observed line width, $\sim$25~km\,s$^{-1}$, is unlike typical maser emission which shows line widths of $\lesssim$1~km\,s$^{-1}$ \citep{patel07}. Given the low $S/N$ of the line, it cannot be ruled out that the profile consists of an ensemble of narrow maser features; indeed water masers are often observed toward Class I objects in the known 22~GHz maser transition \citep[e.g.][]{furuya03}. Only observations with significantly higher sensitivity, such as with ALMA, will be able to exclude this possibility. 

Finally, the two lines may not share a common origin and an isothermal solution is therefore not appropriate. As noted above, the critical density of the 321~GHz line is significantly lower than that of the 63~$\mu$m line by six orders of magnitude. It is therefore possible that the 321~GHz line is predominantly excited in a hot low-density environment, whereas the 63~$\mu$m line requires a denser, possibly cooler, environment. On the other hand, the $^{13}$CO line, with a critical density two orders of magnitude higher than that of the 63~$\mu$m water line, shows a similar line profile to the 321~GHz line, suggesting that these two originate in the same medium. This further suggests that density is not the critical parameter in exciting these transitions, but rather temperature which argues in favor of an isothermal solution. 

The $^{13}$CO emission is interpreted as originating in a disk wind which is unresolved at 0\farcs2 resolution (30 AU) \citep{herczeg11}. The $^{13}$CO column density is not measured toward HL Tau because of the relatively low $S/N$ of the individual rovibrational lines, but similar sources show total $^{12}$CO column densities of $\sim$10$^{19}$~cm$^{-2}$ over emitting radii of 1--4~AU. If the vibrationally excited CO originates in the same component as H$_2$O, this sets an upper limit on the H$_2$O column density. In the case that all oxygen not locked up in CO is driven into H$_2$O, the H$_2$O column density is at most 1.3$\times$10$^{19}$~cm$^{-2}$. This upper limit implies that if the $^{13}$CO and 321~GHz water emission originate in the same physical region, as suggested by the similarity in line profile, then the water is either masing or the simple isothermal non-LTE model is inadequate. An observation of both the 325 GHz para-H$_2^{16}$O line ($J_{KaKc}$=5$_{1,5}$--4$_{2,2}$, $E_{\rm up}$=470~K) and the 321~GHz line with ALMA would be able to break this degeneracy.  The lower-exited line has a critical density similar to the 321~GHz line ($\sim$10$^5$~cm$^{-3}$) but obviously a significantly lower upper-level energy and so the ratio of these two line strengths would probe the temperature dependence of the excitation.

\section{Discussion}

\subsection{Origin of emission}

The PACS water emission from Class II sources has previously been interpreted as originating in the inner hot disk ($\sim$1--3~AU) where the temperature exceeds 600~K \citep{riviere12, howard13}, mainly due to a lack of spatial extent in the PACS data. Emission in these observations is always confined to the central 9\farcs4$\times$9\farcs4 PACS spaxel (typically $\sim$1300~AU), and is only seen toward sources with known jets or winds. 

The observed line profile of the 321~GHz emission shows an offset of 20~km\,s$^{-1}$ from the source velocity. If such an offset is caused by Keplerian rotation in the inner disk at a radius of 1--3~AU, then the stellar mass would be $\sim$0.5--1.5~$M_\odot$. This stellar mass is consistent with what has been inferred previously \citep{sargent91}. However, if the emission originates in a disk, the line profile should be symmetric unless something is hiding the red-shifted counterpart. The disk inclination with respect to the line of sight is estimated to be $\sim$40$^\circ$ \citep{kwon11, alma15} and so the red-shifted counterpart should be visible. Even in the very inner disk where the dust is optically thick at sub-mm wavelengths \citep{alma15}, the surface layers of the entire disk are visible and not just the blue-shifted side. The velocity offset of the line implies an origin outside of the disk. 

Similar line profiles have been observed toward other sources, in particular the more embedded Class 0 sources \citep{kristensen13}, in emission from H$_2$O and light hydrides. The \textit{Herschel}-HIFI observations are interpreted as originating in dissociative wind shocks close to the protostar, where the dissociation is caused by external UV radiation. HL Tau drives a molecular outflow \citep[e.g.][]{cabrit96, lumbreras14, alma15} as well as an optical jet \citep[e.g.][]{movsessian07} and so it is likely that the observed H$_2$O emission is related to one of these components. Indeed, wind models by \citet{panoglou12} and \citet{yvart16} predict that a significant fraction of Class I/II winds are molecular. The data presented here do not have high enough $S/N$ to constrain the type of wind \cite[see ][for a description of different wind types, as well as their diagnostics]{agra-amboage14}.

If the emission is caused by a protostellar wind, the red-shifted lobe of the wind still needs to be hidden. A possibility is that the disk opacity is so high at 321~GHz that the other side is hidden. The recent ALMA high-resolution data \citep{alma15} show that this is indeed the case, at least for the inner $\sim$10~AU or so but likely extending out to and including the first dust ring at a radius of $\sim$20~AU. At larger distances the dust is optically thin, with the exception of a more distant ring at $r$ $\sim$80~AU. These radii put natural limits on the extent of the emitting region which must be smaller than the optically thick dust, i.e. $\lesssim$20~AU.

\subsection{Implications for hot water in protostellar systems}

At most 10\% of the detected 321~GHz water emission toward HL Tau can be assigned to a disk origin, a number which is based on the upper limit of detectable emission centered at the source velocity and for an assumed linewidth of 10~km\,s$^{-1}$. HL Tau is intermediate in evolutionary stage between Class I and II YSOs.  Therefore, the potential implications for interpretation of \textit{Spitzer} and \textit{Herschel} water detections from more evolved T Tauri stars is important and, as yet, unclear. Furthermore, this source drives a strong outflow and jet. Thus, it is possible that HL Tau represents an earlier stage in the evolution where the water in the inner disk is less abundant, and water is predominantly associated with the outflow whereas more evolved sources do indeed have water-rich inner disks. 

Future observations at higher sensitivity and angular resolution will be required to pin down the origin of highly excited water emission more accurately. However, these observations already open a new window for studying hot water from the ground, particularly where the water emission is coming from (inner disk vs. wind) and thus ultimately where water is located in protostellar systems and how much there is of it.

\acknowledgments

The Submillimeter Array is a joint project between the Smithsonian Astrophysical Observatory and the Academia Sinica Institute of Astronomy and Astrophysics and is funded by the Smithsonian Institution and the Academia Sinica. The authors would like to thank J. Najita and J. Carr for fruitful discussions on the origin of hot water emission, and M. Gurwell for assistance with the data calibration. 

{\it Facilities:} \facility{SubMillimeter Array (SMA)}


\begin{thebibliography}{}
\expandafter\ifx\csname natexlab\endcsname\relax\def\natexlab#1{#1}\fi

\bibitem[{{Agra-Amboage} {et~al.}(2014)}]{agra-amboage14} Agra-Amboage, V., Cabrit, S., Dougados, C., et al.\ 2014, \aap, 564, A11 

\bibitem[{{ALMA Partnership} {et~al.}(2015){ALMA Partnership}, {Brogan},
  {P{\'e}rez}, {Hunter}, {Dent}, {Hales}, {Hills}, {Corder}, {Fomalont},
  {Vlahakis}, {Asaki}, {Barkats}, {Hirota}, {Hodge}, {Impellizzeri}, {Kneissl},
  {Liuzzo}, {Lucas}, {Marcelino}, {Matsushita}, {Nakanishi}, {Phillips},
  {Richards}, {Toledo}, {Aladro}, {Broguiere}, {Cortes}, {Cortes}, {Espada},
  {Galarza}, {Garcia-Appadoo}, {Guzman-Ramirez}, {Humphreys}, {Jung}, {Kameno},
  {Laing}, {Leon}, {Marconi}, {Mignano}, {Nikolic}, {Nyman}, {Radiszcz},
  {Remijan}, {Rod{\'o}n}, {Sawada}, {Takahashi}, {Tilanus}, {Vila Vilaro},
  {Watson}, {Wiklind}, {Akiyama}, {Chapillon}, {de Gregorio-Monsalvo}, {Di
  Francesco}, {Gueth}, {Kawamura}, {Lee}, {Nguyen Luong}, {Mangum}, {Pietu},
  {Sanhueza}, {Saigo}, {Takakuwa}, {Ubach}, {van Kempen}, {Wootten},
  {Castro-Carrizo}, {Francke}, {Gallardo}, {Garcia}, {Gonzalez}, {Hill},
  {Kaminski}, {Kurono}, {Liu}, {Lopez}, {Morales}, {Plarre}, {Schieven},
  {Testi}, {Videla}, {Villard}, {Andreani}, {Hibbard}, \& {Tatematsu}}]{alma15}
{ALMA Partnership}, {Brogan}, C.~L., {P{\'e}rez}, L.~M., {et~al.} 2015, \apjl,
  808, L3

\bibitem[{{Bethell} \& {Bergin}(2009)}]{bethell09}
{Bethell}, T., \& {Bergin}, E. 2009, Science, 326, 1675

\bibitem[{{Cabrit} {et~al.}(1996){Cabrit}, {Guilloteau}, {Andre}, {Bertout},
  {Montmerle}, \& {Schuster}}]{cabrit96}
{Cabrit}, S., {Guilloteau}, S., {Andre}, P., {et~al.} 1996, \aap, 305, 527

\bibitem[{{Carr} \& {Najita}(2008)}]{carr08}
{Carr}, J.~S., \& {Najita}, J.~R. 2008, Science, 319, 1504

\bibitem[{{Carr \& Najita}(2011)}]{carr11} Carr, J.~S., \& Najita, J.~R.\ 2011, \apj, 733, 102 

\bibitem[{{Carrasco-Gonz{\'a}lez} {et~al.}(2009){Carrasco-Gonz{\'a}lez},
  {Rodr{\'{\i}}guez}, {Anglada}, \& {Curiel}}]{carrasco09}
{Carrasco-Gonz{\'a}lez}, C., {Rodr{\'{\i}}guez}, L.~F., {Anglada}, G., \&
  {Curiel}, S. 2009, \apjl, 693, L86

\bibitem[{{Daniel} {et~al.}(2011){Daniel}, {Dubernet}, \&
  {Grosjean}}]{daniel11}
{Daniel}, F., {Dubernet}, M.-L., \& {Grosjean}, A. 2011, \aap, 536, A76

\bibitem[{{van Dishoeck} {et~al.}(2014)}]{vandishoeck14} van Dishoeck, 
E.~F., Bergin, E.~A., Lis, D.~C., 
\& Lunine, J.~I.\ 2014, Protostars and Planets VI, 835 

\bibitem[{{Faure} \& {Josselin}(2008)}]{faure08}
{Faure}, A., \& {Josselin}, E. 2008, \aap, 492, 257

\bibitem[{{Fedele} {et~al.}(2013){Fedele}, {Bruderer}, {van Dishoeck}, {Carr},
  {Herczeg}, {Salyk}, {Evans}, {Bouwman}, {Meeus}, {Henning}, {Green},
  {Najita}, \& {G{\"u}del}}]{fedele13}
{Fedele}, D., {Bruderer}, S., {van Dishoeck}, E.~F., {et~al.} 2013, \aap, 559,
  A77

\bibitem[{{Furuya} {et~al.}(2003)}]{furuya03} Furuya, R.~S., Kitamura, 
Y., Wootten, A., Claussen, M.~J., \& Kawabe, R.\ 2003, \apjs, 144, 71 

\bibitem[{{Gallagher} {et~al.}(2003){Gallagher}, {Irace}, \&
  {Werner}}]{gallagher03}
{Gallagher}, D.~B., {Irace}, W.~R., \& {Werner}, M.~W. 2003, in \procspie, Vol.
  4850, IR Space Telescopes and Instruments, ed. J.~C. {Mather}, 17--29

\bibitem[{{Herczeg} {et~al.}(2011){Herczeg}, {Brown}, {van Dishoeck}, \&
  {Pontoppidan}}]{herczeg11}
{Herczeg}, G.~J., {Brown}, J.~M., {van Dishoeck}, E.~F., \& {Pontoppidan},
  K.~M. 2011, \aap, 533, A112

\bibitem[{{Herczeg} {et~al.}(2012){Herczeg}, {Karska}, {Bruderer},
  {Kristensen}, {van Dishoeck}, {J{\o}rgensen}, {Visser}, {Wampfler}, {Bergin},
  {Y{\i}ld{\i}z}, {Pontoppidan}, \& {Gracia-Carpio}}]{herczeg12}
{Herczeg}, G.~J., {Karska}, A., {Bruderer}, S., {et~al.} 2012, \aap, 540, A84

\bibitem[{{Hirota} {et~al.}(2014){Hirota}, {Kim}, {Kurono}, \&
  {Honma}}]{hirota14}
{Hirota}, T., {Kim}, M.~K., {Kurono}, Y., \& {Honma}, M. 2014, \apjl, 782, L28

\bibitem[{{Hogerheijde} {et~al.}(2011)}]{hogerheijde11} Hogerheijde, M.~R., 
Bergin, E.~A., Brinch, C., et al.\ 2011, Science, 334, 338 

\bibitem[{{Howard} {et~al.}(2013){Howard}, {Sandell}, {Vacca}, {Duch{\^e}ne},
  {Mathews}, {Augereau}, {Barrado}, {Dent}, {Eiroa}, {Grady}, {Kamp}, {Meeus},
  {M{\'e}nard}, {Pinte}, {Podio}, {Riviere-Marichalar}, {Roberge}, {Thi},
  {Vicente}, \& {Williams}}]{howard13}
{Howard}, C.~D., {Sandell}, G., {Vacca}, W.~D., {et~al.} 2013, \apj, 776, 21

\bibitem[{{Karska} {et~al.}(2013){Karska}, {Herczeg}, {van Dishoeck},
  {Wampfler}, {Kristensen}, {Goicoechea}, {Visser}, {Nisini}, {San
  Jos{\'e}-Garc{\'{\i}}a}, {Bruderer}, {{\'S}niady}, {Doty}, {Fedele},
  {Y{\i}ld{\i}z}, {Benz}, {Bergin}, {Caselli}, {Herpin}, {Hogerheijde},
  {Johnstone}, {J{\o}rgensen}, {Liseau}, {Tafalla}, {van der Tak}, \&
  {Wyrowski}}]{karska13}
{Karska}, A., {Herczeg}, G.~J., {van Dishoeck}, E.~F., {et~al.} 2013, \aap,
  552, A141

\bibitem[{{Karska} {et~al.}(2014){Karska}, {Kristensen}, {van Dishoeck},
  {Drozdovskaya}, {Mottram}, {Herczeg}, {Bruderer}, {Cabrit}, {Evans},
  {Fedele}, {Gusdorf}, {J{\o}rgensen}, {Kaufman}, {Melnick}, {Neufeld},
  {Nisini}, {Santangelo}, {Tafalla}, \& {Wampfler}}]{karska14}
{Karska}, A., {Kristensen}, L.~E., {van Dishoeck}, E.~F., {et~al.} 2014, \aap,
  572, A9

\bibitem[{{Kristensen} {et~al.}(2013){Kristensen}, {van Dishoeck}, {Benz},
  {Bruderer}, {Visser}, \& {Wampfler}}]{kristensen13}
{Kristensen}, L.~E., {van Dishoeck}, E.~F., {Benz}, A.~O., {et~al.} 2013, \aap,
  557, A23

\bibitem[{{Kwon} {et~al.}(2011){Kwon}, {Looney}, \& {Mundy}}]{kwon11}
{Kwon}, W., {Looney}, L.~W., \& {Mundy}, L.~G. 2011, \apj, 741, 3

\bibitem[{{Lumbreras} \& {Zapata}(2014)}]{lumbreras14}
{Lumbreras}, A.~M., \& {Zapata}, L.~A. 2014, \aj, 147, 72

\bibitem[{{McMullin} {et~al.}(2007){McMullin}, {Waters}, {Schiebel}, {Young},
  \& {Golap}}]{mcmullan07}
{McMullin}, J.~P., {Waters}, B., {Schiebel}, D., {Young}, W., \& {Golap}, K.
  2007, in Astronomical Data Analysis Software and Systems XVI (ASP Conf. Ser.
  376), ed. R.~A. {Shaw}, F.~{Hill}, \& D.~J. {Bell}, 127

\bibitem[{{Menten} \& {Melnick}(1991)}]{menten91}
{Menten}, K.~M., \& {Melnick}, G.~J. 1991, \apj, 377, 647

\bibitem[{{Menten} {et~al.}(1990){Menten}, {Melnick}, \& {Phillips}}]{menten90}
{Menten}, K.~M., {Melnick}, G.~J., \& {Phillips}, T.~G. 1990, \apjl, 350, L41

\bibitem[{{Movsessian} {et~al.}(2007){Movsessian}, {Magakian}, {Bally},
  {Smith}, {Moiseev}, \& {Dodonov}}]{movsessian07}
{Movsessian}, T.~A., {Magakian}, T.~Y., {Bally}, J., {et~al.} 2007, \aap, 470,
  605

\bibitem[{{Neufeld} \& {Melnick}(1990)}]{neufeld90}
{Neufeld}, D.~A., \& {Melnick}, G.~J. 1990, \apjl, 352, L9

\bibitem[{{Panoglou} {et~al.}(2012){Panoglou}, {Cabrit}, {Pineau Des
  For{\^e}ts}, {Garcia}, {Ferreira}, \& {Casse}}]{panoglou12}
{Panoglou}, D., {Cabrit}, S., {Pineau Des For{\^e}ts}, G., {et~al.} 2012, \aap,
  538, A2

\bibitem[{{Patel} {et~al.}(2007){Patel}, {Curiel}, {Zhang}, {Sridharan}, {Ho},
  \& {Torrelles}}]{patel07}
{Patel}, N.~A., {Curiel}, S., {Zhang}, Q., {et~al.} 2007, \apjl, 658, L55

\bibitem[{{Pilbratt} {et~al.}(2010){Pilbratt}, {Riedinger}, {Passvogel},
  {Crone}, {Doyle}, {Gageur}, {Heras}, {Jewell}, {Metcalfe}, {Ott}, \&
  {Schmidt}}]{pilbratt10}
{Pilbratt}, G.~L., {Riedinger}, J.~R., {Passvogel}, T., {et~al.} 2010, \aap,
  518, L1

\bibitem[{{Pontoppidan} {et~al.}(2010){Pontoppidan}, {Salyk}, {Blake}, \&
  {K{\"a}ufl}}]{pontoppidan10a}
{Pontoppidan}, K.~M., {Salyk}, C., {Blake}, G.~A., \& {K{\"a}ufl}, H.~U. 2010,
  \apjl, 722, L173

\bibitem[{{Riviere-Marichalar} {et~al.}(2012){Riviere-Marichalar},
  {M{\'e}nard}, {Thi}, {Kamp}, {Montesinos}, {Meeus}, {Woitke}, {Howard},
  {Sandell}, {Podio}, {Dent}, {Mendigut{\'{\i}}a}, {Pinte}, {White}, \&
  {Barrado}}]{riviere12}
{Riviere-Marichalar}, P., {M{\'e}nard}, F., {Thi}, W.~F., {et~al.} 2012, \aap,
  538, L3

\bibitem[{{Salyk} {et~al.}(2008){Salyk}, {Pontoppidan}, {Blake}, {Lahuis}, {van
  Dishoeck}, \& {Evans}}]{salyk08}
{Salyk}, C., {Pontoppidan}, K.~M., {Blake}, G.~A., {et~al.} 2008, \apjl, 676,
  L49

\bibitem[{{Salyk} {et~al.}(2011)}]{salyk11} Salyk, C., Pontoppidan, K.~M., Blake, G.~A., Najita, J.~R., \& Carr, J.~S.\ 2011, \apj, 731, 130 

\bibitem[{{Sargent} \& {Beckwith}(1991)}]{sargent91}
{Sargent}, A.~I., \& {Beckwith}, S.~V.~W. 1991, \apjl, 382, L31

\bibitem[{{Stevenson} \& {Lunine}(1988)}]{stevenson88}
{Stevenson}, D.~J., \& {Lunine}, J.~I. 1988, Icarus, 75, 146

\bibitem[{{Torres} {et~al.}(2009){Torres}, {Loinard}, {Mioduszewski}, \&
  {Rodr{\'{\i}}guez}}]{torres09}
{Torres}, R.~M., {Loinard}, L., {Mioduszewski}, A.~J., \& {Rodr{\'{\i}}guez},
  L.~F. 2009, \apj, 698, 242

\bibitem[{{van der Tak} {et~al.}(2007){van der Tak}, {Black}, {Sch{\"o}ier},
  {Jansen}, \& {van Dishoeck}}]{vandertak07}
{van der Tak}, F.~F.~S., {Black}, J.~H., {Sch{\"o}ier}, F.~L., {Jansen}, D.~J.,
  \& {van Dishoeck}, E.~F. 2007, \aap, 468, 627

\bibitem[{{Woitke} {et~al.}(2009)}]{woitke09} Woitke, P., Kamp, I., \& Thi, W.-F.\ 2009, \aap, 501, 383 

\bibitem[{{Yvart} {et~al.}(2016){Yvart}, {Cabrit}, {Pineau des For{\^e}ts}, \&
  {Ferreira}}]{yvart16}
{Yvart}, W., {Cabrit}, S., {Pineau des For{\^e}ts}, G., \& {Ferreira}, J. 2016,
  \aap, 585, A74

\end{thebibliography}
\end{document}